\documentclass[a4paper,12pt]{article}
\usepackage[utf8]{inputenc}
\usepackage[T1,T2A]{fontenc}
\usepackage{amsmath,amsfonts,amssymb}
\usepackage{cite}

\begin{document}

\renewcommand*{\thefootnote}{\fnsymbol{footnote}}

\begin{center}
{\Large\bf Motivations for the Soft Wall holographic approach to strong interactions
}
\end{center}
\vspace{-0.1cm}
\begin{center}
{Sergey Afonin\footnote{E-mail: \texttt{s.afonin@spbu.ru}},
Timofey Solomko
}
\end{center}

\renewcommand*{\thefootnote}{\arabic{footnote}}
\setcounter{footnote}{0}

\begin{center}
{\small Saint Petersburg State University, 7/9 Universitetskaya nab.,
St.Petersburg, 199034, Russia}
\end{center}

\bigskip

{\footnotesize
\noindent
The hypothesis of gauge/gravity correspondence (or holographic duality) from string
theory has led to unexpected and challenging ways
for description of strongly coupled systems. Such descriptions are
given in terms of weakly-coupled higher-dimensional gravitational
theories. The holographic ideas have penetrated many branches
of theoretical physics. We discuss motivations for the so-called Soft Wall holographic approach to
non-perturbative strong interactions and its advantages.
}

\bigskip
\bigskip

The bottom-up holographic approach to strong interactions encompasses now a large number of phenomenological models inspired by the gauge/gravity duality in string theory~\cite{mald,witten,gub}. In this approach, one applies the holographic methods developed for conformal field theories to the case of real QCD.
The holographic approach is based on a conjecture that observables in strongly coupled gauge theories, in the limit of a large number of colors, can be determined from classical fields  weakly coupled through gravity in Anti-de Sitter (AdS) space that has one extra space dimension. The symmetry principles play a crucial role
in this conjecture.

The bottom-up holographic approach was formulated in its final form in papers~\cite{son1,pom,Hirn:2005nr} where it was applied to description of
spontaneous chiral symmetry breaking and spectrum of light mesons.
The incorporation of Chern-Simons term allowed to describe baryons and the physics related to the QCD chiral anomaly~\cite{Pomarol:2008aa}.
The overall agreement of the bottom-up holographic approach with the existing hadron phenomenology turned out to be surprisingly good.
This marked the birth of a new class of models describing the low-energy QCD, hadron spectra, hadron structure, and QCD thermodynamics 
with an accuracy comparable with old traditional approaches (effective field theories, potential quark models, etc.) and with a similar number of
input parameters. Since that time (2005)
a great number of various bottom-up holographic models for strongly coupled QCD have been proposed and applied to description of the hadron phenomenology.

In the bottom-up holographic approach, following the ideas of gauge/gravity duality (namely its realization in the AdS/CFT correspondence~\cite{mald,witten,gub}), 
one assumes the existence of 5D dual theory (in the sense of strong-weak duality) for QCD in the large-$N_c$ limit and tries to build a phenomenologically
useful gravitational 5D model. This putative 5D theory, as in the AdS/CFT correspondence, is constructed
in the AdS$_5$ space or asymptotically AdS$_5$ space. This ensures the correct high-energy asymptotics of correlation functions in QCD
which follow the pattern of conformal field theory due to the asymptotic freedom.
From the geometrical viewpoint,
the 4D boundary of AdS$_5$ is time-like (and only in the case of this space if we wish to have a homogeneous space of constant curvature)
and has the appearance (at least after appropriate choice
of coordinates) of 4D Minkowski flat space. Also the AdS$_5$ space emerges naturally if one tries to ``geometrize''  the dilatation symmetry in a 4D conformal theory,
i.e., to rewrite the scaling transformations of field theory operators as space-time transformations of some 5D fields~\cite{sundrum}. For more specific motivations
which are relevant to QCD see Ref.~\cite{pol}. 

A convenient parametrization of the AdS$_5$ metric is given by the Poincar\'{e} patch with the line element
\begin{equation}
ds^2=\frac{R^2}{z^2}\left(\eta^{\mu\nu}dx_\mu dx_\nu-dz^2\right),
\end{equation}
where $R$ is the radius of AdS$_5$ space and $z\geq0$ represents the fifth holographic coordinate that has
the physical meaning of inverse energy scale. The 4D Minkowski space becomes the ultraviolet boundary
of AdS$_5$ residing at $z=0$. The relation between a 4D gauge theory and its dual 5D gravitational theory
schematically is given by a concise statement
\begin{equation}
S_\text{eff}\{\phi(x)\}+J(x)\phi(x)=\left.S_\text{5D}^\text{boundary}\left(\phi(x,z)\right)\right|_{\phi(x,0)\doteq J(x)},
\end{equation}
where $\phi(x,0)$ represents the UV boundary value of $\phi(x,z)$.
If the 4D gauge theory is in the strong coupling regime, the 5D theory must be weakly coupled due to
strong-weak duality. This general idea paved the way for building semiclassical 5D models
which describe the low energy QCD and are often extended to higher energies.

The holographic approach to strong interactions includes also various top-down
holographic models which start from some brane construction within a string theory and try to get a dual model useful for the QCD phenomenology. A natural implementation of Regge behavior in the hadron spectra and correct Operator Product Expansion (OPE) of correlation functions in QCD have still not been achieved in the top-down approach.

Many bottom-up holographic models were built on the base
of the so-called Soft Wall (SW) holographic model introduced in~\cite{son2,andreev}. The mass scale $|\lambda|$ is incorporated into these models via the exponential
scale factor $e^{\lambda z^2}$ in 5D action of the dual theory~\cite{son2}. Here $z$ is the fifth coordinate called holographic which is interpreted as the inverse energy scale.
The given scale factor is often called ``dilaton background'' or just
``background''. This ``background'' should have a dynamical origin,
in particular, it was suggested to be a result of a closed string tachyon condensation in the original paper~\cite{son2}.
We are not aware of any explicit realization of this proposal but a bottom-up holographic model based on an open string tachyon condensation
(adopted in a simplified form from the string theory) was constructed in Ref.~\cite{Casero:2007ae} and worked out further in
Refs.~\cite{Gursoy:2007cb,Gursoy:2007er,Iatrakis:2010jb}.
Remarkably, the setup introduced in~\cite{Casero:2007ae} describes both the chiral symmetry breaking and asymptotically linear radial Regge trajectories.

The standard SW holographic model is defined by the 5D action~\cite{son2}
\begin{equation}
  S=\int d^4\!x\,dz\sqrt{g}\,e^{\lambda z^2}\mathcal{L},
\end{equation}
where $g=|\text{det}g_{MN}|$, $\mathcal{L}$ is a Lagrangian density of some
free fields in AdS\(_5\) space which, by assumption, are
dual on the AdS\(_5\) boundary to some QCD operators.
The 4D mass spectrum of this model can be found
from the equation of motion accepting the 4D plane-wave ansatz. For instance, in the case
of free massless vector fields, one considers the particle-like ansatz
$V_\mu(x,z)=e^{ipx}v(z)\epsilon_\mu$ with the on-shell, $p^2=m^2$, and transversality,
$p^\mu\epsilon_\mu=0$, conditions~\cite{son2}. The resulting spectrum takes the Regge form,
\begin{equation}
\label{8}
m_n^2=4|\lambda|n,\qquad n=1,2,\dots.
\end{equation}
This model can be generalized to the case of arbitrary intercept $b$ in the spectrum~\cite{Afonin:2012jn,Afonin:2021cwo},
$m_n^2=4|\lambda|(n+b)$.

The original SW model was designed to
describe the phenomenology of linear Regge trajectories in the large-$N_c$ limit of QCD but it turned out successful in other areas of hadron phenomenology,
in many cases demonstrating an intelligible interpolation between the low and high energy sectors of QCD.
The simplest SW holographic model can be viewed as the most self-consistent way of rewriting the infinite number of pole terms
(expected in the large-$N_c$ limit of QCD) with linear spectrum of masses squared
in the pole representation of two-point Correlation Functions (CFs), as some 5D gravitational model of free fields~\cite{afonin2010,ahep}.
Remarkably, the holographic recipe of Refs.~\cite{witten,gub} for calculation of CFs follows in a natural way within such a rewriting.
This means, in particular, that the SW holographic models are closely related
with the planar QCD sum rules (in a sense, they represent just 5D rewriting of those sum rules~\cite{afonin2010})
which were widely used in the past to study the phenomenology of linear radial trajectories in the meson sector(see, e.g.,~\cite{lin2}).
In descriptions of hadron electromagnetic form factors, the holographic approach, especially the SW one,
recovers the old pre-QCD dual description with all its phenomenological successes~\cite{zahed3}.
But the holographic QCD is much wider in scope --- its strong advantage consists in the use of
Lagrangian formulation that enables more refined calculations and opens the door to many other applications.

A top-down derivation of SW like holographic models from some brane construction in a string theory remains an open problem.
This problem, however, is purely theoretical and does not represent an obstacle for building a rich holographic phenomenology.
It should be recalled that a similar situation persists in non-perturbative QCD --- there are various popular phenomenological models
for low-energy strong interactions (various sigma-models, the Nambu--Jona-Lasinio model, etc.)
but no one of them has been derived from QCD. We believe that the bottom-up holographic models will remain useful and actively explored even if
in future somebody proves rigorously that the gauge/gravity duality cannot work for non-conformal theories. Such a result would make outdated
the holographic top-down approach and likely some bottom-up models but the SW (and likely some elements of Hard Wall) approach would
survive because, as we mentioned above, it became a useful language unifying into one logical framework many elements of various old
phenomenological approaches (QCD sum rules in the large-$N_c$ limit, light-cone QCD, Regge physics, deep inelastic scattering, chiral perturbation theory)
and reproducing many results from those approaches. For this reason, the further development of this 5D language
for hadron physics looks perspective.

The numerous practical applications show usually that the approximation of a static dilaton background and of probe limit
(i.e., when the 5D metrics is not back-reacted neither by a dilaton background nor by the matter fields)
is more than enough for phenomenological purposes.
Furthermore, we are not aware of any bottom-up holographic model beyond this approximation or a top-down holographic model for QCD
that would reproduce correctly the analytical structure of OPE of correlation functions in QCD (i.e., the perturbative logarithm plus power corrections).
It should be stressed
that the bottom-up AdS/QCD is a phenomenology-driven approach, so the agreement with the known phenomenology should be in first place in any judgement
about ``correctness'' of a model.

From a conceptual viewpoint, the consistency of dynamical holographic models with back-reaction is questionable
if the whole approach somehow follows from an underlying string theory.
Indeed, both the gravitational metric and dilaton background are then determined by underlying string dynamics,
hence, the metric is back-reacted by dilaton (and vice versa) indirectly, via this string dynamics, i.e., the given back-reaction cannot be fully
described just by a set of coupled Einstein equations for the metric and dilaton. An instructive example of this point is given
by an extensive analysis of Refs.~\cite{Casero:2007ae,Gursoy:2007cb,Gursoy:2007er,Iatrakis:2010jb},
where it was advocated that in exploring improved holographic theories for QCD
a seminal direction is to think of the 5D bulk theory as a (non-critical) string theory, not just gravity. It turns out that
it is a gaussian potential of scalar tachyon field that can give rise to a SW like background.
We see thus that the assumption that the underlying string dynamics can be
neglected in such a way that some effective dynamical dilaton persists, as a matter of fact, looks almost as poorly substantiated theoretically
as the assumption of effective static dilaton background. As long as the underlying string dynamics is unknown,
both the dynamic and static dilaton background represent just working hypotheses for building bottom-up models. Only the phenomenology can discriminate
which hypothesis works better, hence, is more ``right''. To the best of our knowledge, the most successful holographic model describing the Regge physics,
OPE of CFs, and hadron form-factors is the static SW holographic model. There is no purely theoretical justification
for this observation. In the pioneering paper~\cite{son2}, it was assumed that it is a closed string tachyon
condensation that should lead to a static dilaton background. The form of this background is dictated by the QCD phenomenology.

In summary, among the practitioners of SW holographic models there is
a widespread belief that these models efficiently interpolate
between low energy and high energy QCD. In certain sense, they can be seen as a ``meromorphization'' of the perturbative QCD
expression for two-point correlation functions~\cite{br}. Many references on interesting applications of SW holographic approach
can be found in~\cite{Afonin:2021cwo}.

\underline{\bf Acknowledgements.}
This research was funded by the Russian Science Foundation grant number 21-12-00020.

\end{document}